\documentclass[runningheads]{llncs}
\usepackage[T1]{fontenc}
\usepackage{graphicx}

\usepackage{url}
\usepackage[hidelinks]{hyperref}
\usepackage[small]{caption}
\usepackage{amsmath}
\usepackage{booktabs}
\usepackage{algorithm}
\usepackage{algorithmic}
\usepackage{csquotes}
\usepackage{subcaption}

\begin{document}

\title{Human-Centered LLM-Agent User Interface: \\A Position Paper}

\author{Daniel Chin\inst{1,2}\orcidID{0000-0002-3406-5318} \and
Yuxuan Wang\inst{1,3}\orcidID{0009-0002-9161-6961} \and
Gus Xia\inst{2}}

\authorrunning{D. Chin et al.}

\institute{
New York University Shanghai 
\email{\{daniel.chin , yw5343\}@nyu.edu}\\
\and
Mohamed bin Zayed University of Artificial Intelligence \email{gus.xia@mbzuai.ac.ae}\\
\and
New York University Tandon School of Engineering
}

\maketitle

\begin{abstract}
Large Language Model (LLM) -in-the-loop applications have been shown to effectively interpret the human user's commands, make plans, and operate external tools/systems accordingly. Still, the operation scope of the LLM agent is limited to passively following the user, requiring the user to frame his/her needs with regard to the underlying tools/systems. We note that the potential of an LLM-Agent User Interface (LAUI) is much greater. A user mostly ignorant to the underlying tools/systems should be able to work with a LAUI to discover an emergent workflow. Contrary to the conventional way of designing an explorable GUI to teach the user a predefined set of ways to use the system, in the ideal LAUI, the LLM agent is initialized to be proficient with the system, proactively studies the user and his/her needs, and proposes new interaction schemes to the user. To illustrate LAUI, we present Flute X GPT, a concrete example using an LLM agent, a prompt manager, and a flute-tutoring multi-modal software-hardware system to facilitate the complex, real-time user experience of learning to play the flute. 

\end{abstract}

\section{Introduction}

\begin{figure}[tb]
  \centering
  \includegraphics[width=0.7\linewidth]{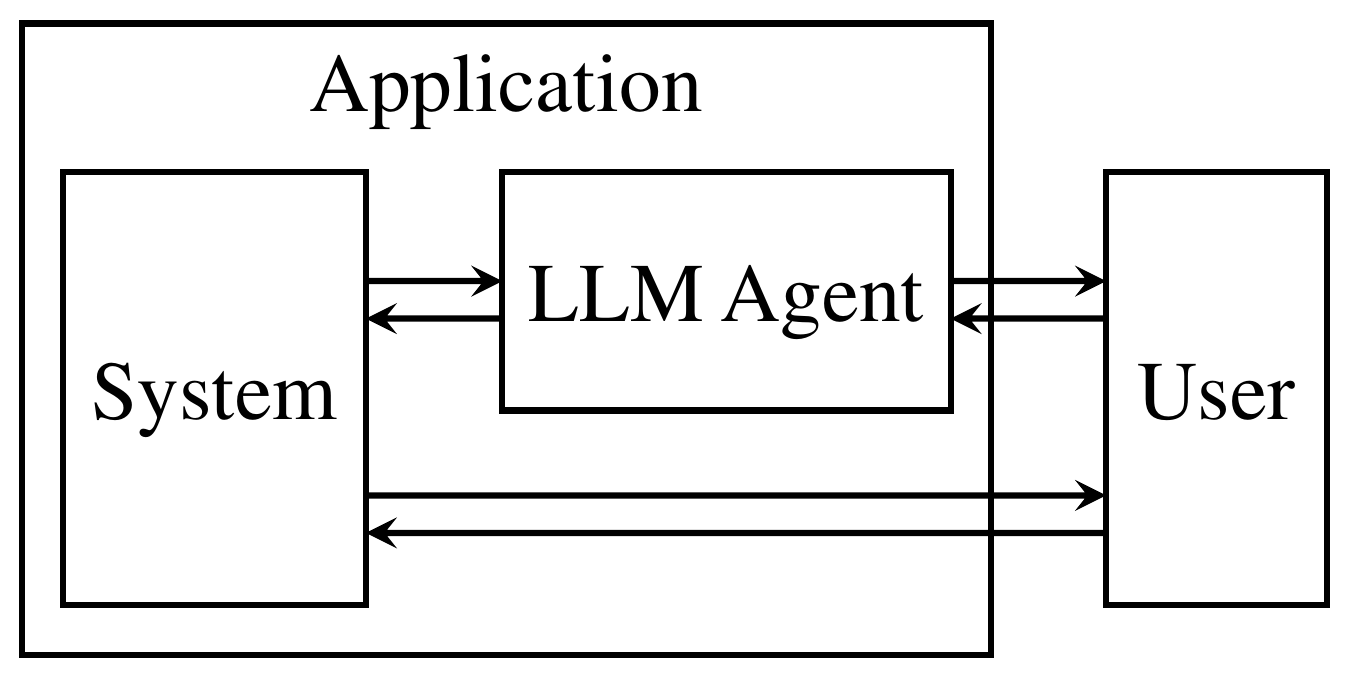}
  \caption{The LLM agent serves as the interface between the underlying system and the user. The LLM agent together with the system forms the application. Direct communications between the user and the system is available and configured by the LLM agent.}
  \label{fig:sys+llma=app}
  \vspace{-0.1in}
\end{figure}

Large Language Models (LLMs) can be used to connect an underlying \textit{system} with a user via the natural language medium, forming an LLM-powered \textit{application} as shown in Figure~\ref{fig:sys+llma=app}. The LLM is embedded in an LLM-in-the-loop state machine to acquire multi-modal input/output capabilities, emulate logical reasoning and planning, and use tools or operate a system 
\cite{shen2024hugginggpt,liu2023controlllm,liang2024taskmatrix,tao2023webwise}. 
Consequently, the user is able to indirectly but effectively use the underlying system via chatting with an LLM assistant. 

However, the current applications hardly address the user-interaction potential stemming from the multi-round chatting setup. Although the user can refer to the chat history, the LLM assistant rarely challenges the user or even asks to clarify the user's intention. Instead, the LLM closely follows the user's command, missing the golden opportunity to improve the user's usage scheme and understanding of the system/tools. That inefficacy becomes jarringly apparent when one tries to design from scratch a new complex system with an LLM interface.

We posit that the operation scope of an LLM-Agent User Interface (LAUI) is much wider than that. The interface should be more than an assistant or a butler, but instead a secretary, actively working with the user to discover emergent interaction schemes on the fly. LAUI should be proficient with the underlying system, study the user, study the user's needs (instead of commands), reason on its own, and propose tailored interaction schemes to the user, including what modes of feedback are provided and what input is expected from the user. In the conventional way of interaction, including GUI and current LLM-powered applications \cite{MicrosoftCopilot}, the designer has to imagine possible usage workflows given the system capabilities at design time, and the user is expected to learn the system (via tutorials, explorations, and practicing) in order to come up with a workflow for each task. In contrast, with LAUI, the user only needs to describe his/her needs and doesn't need to deeply understand the application, and a workflow will naturally emerge as the LLM agent works with the user. We call for more research exploring the potential of LAUI. 

As a concrete illustration of LAUI, we present Flute X GPT --- an LLM-in-the-loop music-tutoring application consisting of an LLM agent, a prompt manager, a software system, and hardware. The application provides real-time haptic guidance via servo motors, real-time visual music-symbol feedback, real-time audio feedback, and natural language chat, all controlled by the LLM agent. This is the first time an LLM-powered interface is applied to a working system of such complexity and real-time user interactivity.

We first describe Flute X GPT in Section~\ref{sec:flute-x-gpt}, illustrating what a specific LAUI can look like, and then go on to formulate the general LAUI in Section~\ref{sec:LAUI}.

\section{Flute X GPT} \label{sec:flute-x-gpt}

We describe Flute X GPT\footnote{The source code is open to public at \url{https://github.com/Daniel-Chin/Flute-X-GPT}}, a music-tutoring application using LAUI. The human user works with Flute X GPT in workshop episodes, practicing to play the flute and learning music. The application gives real-time multi-modal feedback, including haptic feedback that applies force to the user's fingers, visual feedback displaying performance errors, audio feedback rendering the music, and natural-language speech given by a robot music teacher. The underlying software-hardware system has numerous different configurations, each leading to a different interaction workflow. Each setting (e.g., toggle certain feedback, conditions for triggering feedback) can be controlled independently, so the number of configurations grow exponentially with the number of settings. For the user, it is unrealistic to first master the complex system before using it. Even for the designer, the space of possible interaction schemes is impossible to enumerate during design time. The LLM agent steps in to bridge that gap. Via prompting, we instruct the LLM agent to be proficient with all the raw capabilities of the system. During use time, the LLM agent converses with the user to clarify what interaction workflow will benefit the user's music learning goal the most. The LLM agent studies the user's preferences and diagnoses the musical challenges the user is facing. When configuring the underlying system, the LLM agent uses its pretrained common sense to reason about the implied consequences of the system configurations. Certain mixtures of settings have never been previously considered by any human designer, but can still emerge during use time.

\begin{figure}[tb]
    \centering {
    \begin{subfigure}{\linewidth}
        \centering
        \includegraphics[scale=0.8]{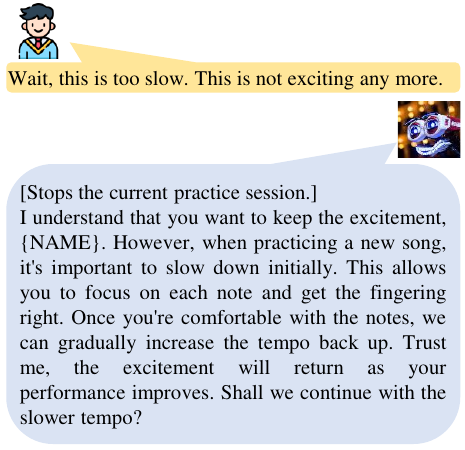}
        \caption{From the scripted trial.}
        \label{fig:dialog-a}
    \end{subfigure}
    \begin{subfigure}{\linewidth}
        \centering
        \includegraphics[scale=0.8]{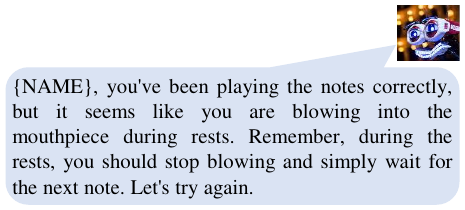}
        \caption{From improvised trial 1.}
        \label{fig:dialog-b}
    \end{subfigure}
    \caption{Interaction excerpts from the video demos.}
    \label{fig:dialog}
    }
    \vspace{-0.1in}
\end{figure}

Overall, Flute X GPT entails three novel contributions:  
\begin{itemize}
    \item \textbf{An illustration of LAUI}. The LLM agent not only follows the user's commands, but also proactively clarifies the user's needs, deduces better interaction workflows, and advises the user. 
    \item \textbf{LLM agent controls real-time system}. The LLM agent directs a real-time interaction that is music training. The LLM agent is aware of time passage and decides when to wait for further event notifications and when to interrupt the user.
    \item \textbf{LLM agent operates complex, stateful, multi-modal, user-interactive system}. The LLM agent operates a highly complex software-hardware system by understanding how the user can benefit from the application and considering what combination of settings will lead to what interactive effects for the user.
\end{itemize}

\subsection{User Experience} \label{subsec:ux}

\begin{table*}[tb]
\caption{Functions that the LLM agent can call to control Music X Machine, the underlying music-tutoring system. The description addresses the LLM agent.}
\centering \tiny {
\begin{tabular}{@{}lll@{}}
\toprule
  \textbf{Function}  & \textbf{Parameters}                                                                           & \textbf{Description}                                                                                                                  \\ \midrule
Wait         & & \begin{tabular}[c]{@{}l@{}} Do nothing and wait for further stimuli, e.g. student speaking/playing music.           \end{tabular}    \\ \midrule
StartSession         & & \begin{tabular}[c]{@{}l@{}}Start a Practice Session on Music X Machine. Do not call this function\\ unless you have already set all the modes. \end{tabular}    \\ \midrule
InterruptSession          && \begin{tabular}[c]{@{}l@{}}Immediately end the Practice Session on Music X Machine. Call when the\\ student is having trouble or has started speaking in the middle of a Session. \end{tabular}                           \\ \midrule
SetHapticMode  & mode                  & \begin{tabular}[c]{@{}l@{}}Set the haptic mode of Music X Machine. \end{tabular}                                                            \\ \midrule
ToggleVisual  & state                  & \begin{tabular}[c]{@{}l@{}}Set the visual KR feedback to be on or off. \end{tabular}                                                            \\ \midrule
PlayReference  &                   & \begin{tabular}[c]{@{}l@{}}Play the ground-truth audio of the current segment.  \end{tabular}                                                            \\ \midrule
LoadSong  & song\_title                  & \begin{tabular}[c]{@{}l@{}}Load a song into Music X Machine, and automatically select the entire \\song as the current segment. It doesn't start a Practice Session by itself.  \end{tabular}                                                            \\ \midrule
SelectSegment  & begin, end                  & \begin{tabular}[c]{@{}l@{}}Select a temporal segment of the song.  \end{tabular}                                                            \\ \midrule
ModifyTempo  & tempo\_multiplier                  & \begin{tabular}[c]{@{}l@{}}Modify the tempo of the song.  \end{tabular}                                                                 \\\bottomrule
\end{tabular}
}
\label{tab:funcs}
\vspace{-0.1in}
\end{table*}

\textbf{The intended user experience}. The user has no prior knowledge about the flute tutoring system. The only assumption about the user is that the user wants to learn the flute. It is the LLM agent's job to adapt to the user's current flute playing capability, other musical skills, demographics, vocabulary, patience, style of learning, etc. During the music learning workshop, the LLM agent wears the face of a robot music teacher to chat with the user. For example, the robot teacher asks the user to put on a pair of haptic gloves, and explains that they provide force feedback at each finger. The workshop then alternates between practice sessions where the user plays a segment of a song under real-time guidance and verbal dialogues between the user and the robot teacher. The user gradually experiences more and more interaction schemes that trains musicality in different ways, and expresses their ideas about using the application and music learning in general. The LLM agent uses that opportunity to study the user and steers the workshop accordingly, aiming to maximize music education effect. The user learns to treat the robot teacher as a considerate and professional music tutoring agent capable of thinking multiple steps ahead, formulating plans with the user, and explaining the plans as well as music knowledge and education principles to the user.

We present three video demos of real-human user tests, including one scripted trial and two improvised trials.\footnote{Demo playlist of three videos: \url{https://www.youtube.com/playlist?list=PLNb0mNThMXbkBPL_Rjtmhx2daxuB6GFLs}}

\textbf{Video demo: Scripted trial}. In this demo, all actors follow a script generated by an offline but faithfully emulated interaction between a user and the LLM agent. To re-emphasize, the LLM agent's lines in the script are not written by humans, but are outputted by the LLM agent. Scripting removes the latency of LLM autoregressive generation. To ensure the script is concise and demonstrates a wide range of behaviors, we intervene with the script generation process. Every turn, we select one response out of 4 to 16 candidates sampled by the LLM. Seldomly, we add a temporary user-role prompt to give a short hint to the LLM agent, or edit the generated response. All the above interventions are kept to the minimum to ensure the vast majority of the agent speech is the authentic output of the LLM. See Figure~\ref{fig:dialog-a} for an excerpt from the video. 

\textbf{Video demo: Two improvised trials}. Flute X GPT is set loose to freely interact with the user. The user is played by a developer pretending to be ignorant to the system. Figure~\ref{fig:dialog-b} shows an excerpt where the developer is surprised by Flute X GPT noticing that he was not playing the rest notes according to the score, a behavior never considered during design time. 

\subsection{System Capabilities} \label{subsec:sys-capabilities}
This subsection lists the capabilities of the music-tutoring system, Music X Machine, that underlies the LLM agent. 

\textbf{Haptic guidance}. The hardware includes a pair of gloves with servo motors and movable finger rings. Through these gloves, the system moves the user's fingers to help with performance motions. The haptic guidance can be configured via various settings (e.g., is the guidance sustained throughout each note or applied at each note onset? Apply guidance for each note? For incorrectly played notes? For unplayed notes?). Certain combinations of settings form configuration \textit{presets} whose interaction scheme is deemed meaningful by the designer. For example, in the Force Mode preset, every note triggers a full-force guidance for each finger. In the Adaptive Mode preset, the user plays the song on his/her own, and the gloves correct the mistakes. The haptic configuration should be tuned to adapt to the user's skill level and the song's difficulty \cite{Zhang2019}.

\textbf{Visual feedback}. A monitor displays the score of the current song. The user-played notes are displayed on top of the notes in the score, yielding the real-time visual Knowledge-of-Result (KR) feedback. It is a visual cue for the user to know where he/she is in terms of the pitch and helps the user internalize the score notations \cite{NIME20_40}. It can be toggled on or off. 

\textbf{Audio feedback}. There are three streams of audio: synthesized user-played flute sounds, teacher-played reference performance audio, and metronomes. The three streams are mixed down and outputted.

\textbf{Sensor-augmented flute}. The flute measures real-time information including finger positions and breath pressure.

\begin{figure*}[tb]
    \centering {
    \begin{subfigure}[b]{0.25\linewidth}
        \includegraphics[width=\linewidth]{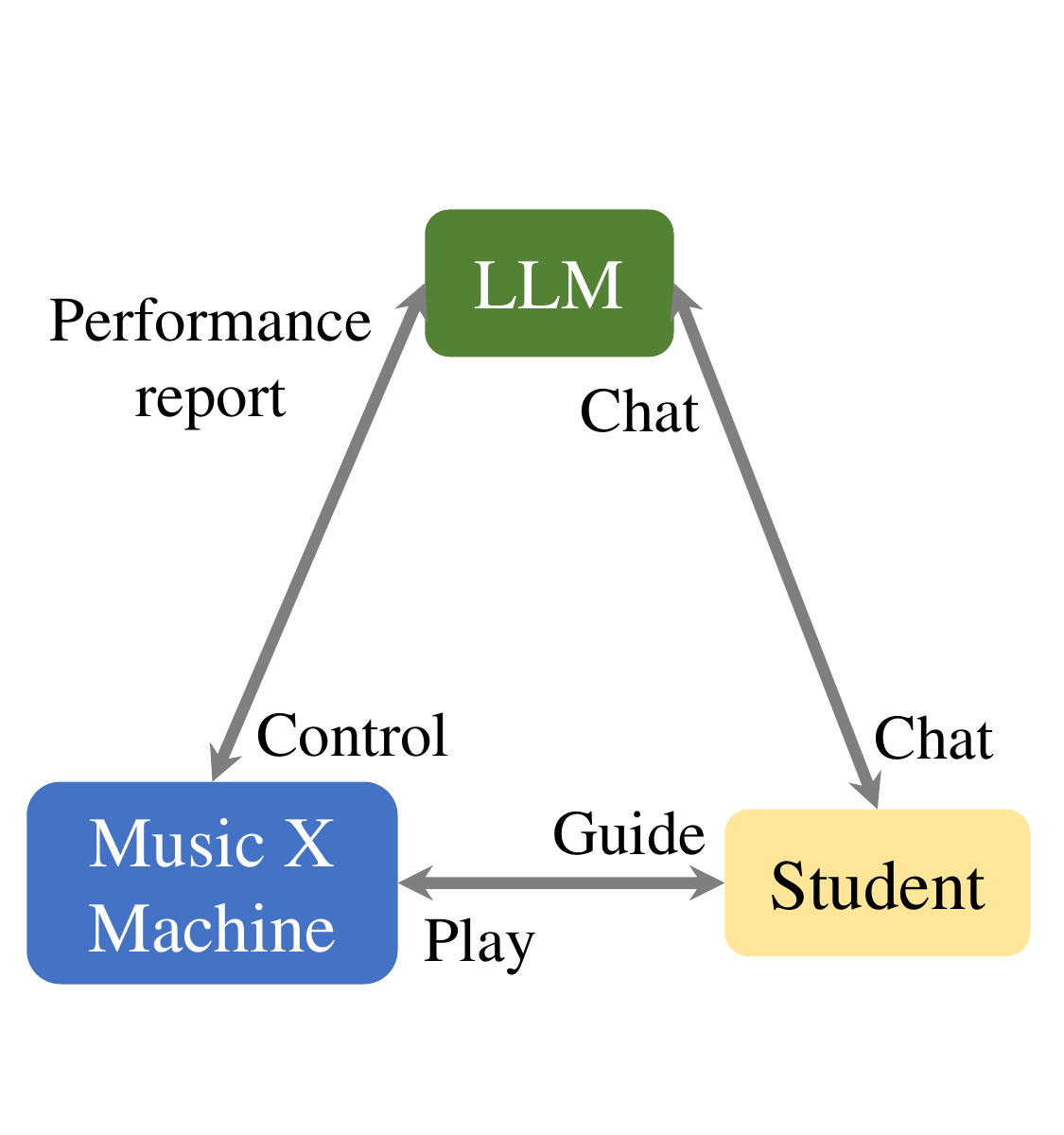}
        \caption{Simplified.}
    \end{subfigure}
    \hspace{0.04\linewidth}
    \begin{subfigure}[b]{0.64\linewidth}
        \includegraphics[width=\linewidth]{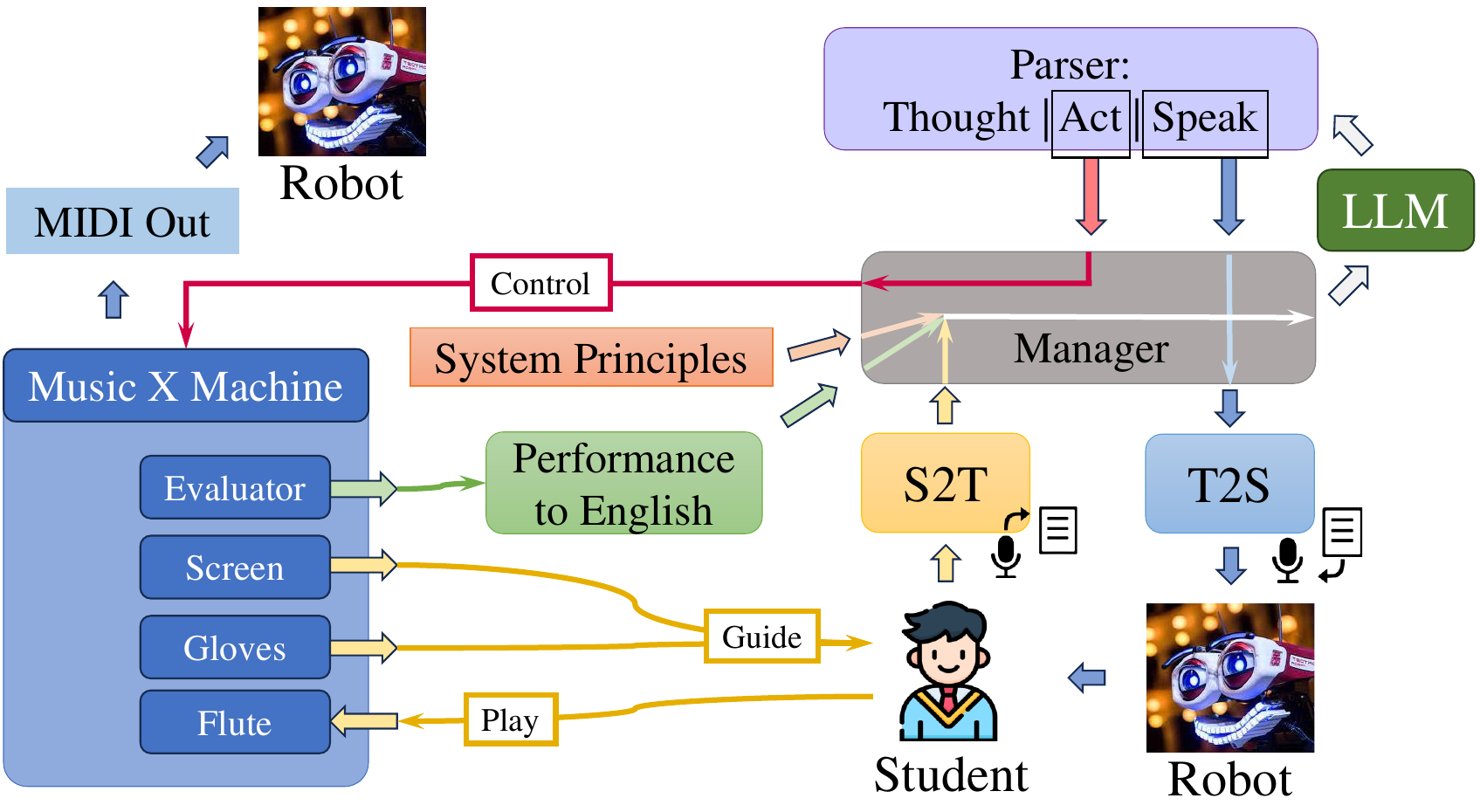}
        \caption{Full.}
    \end{subfigure}
    \caption{Flute X GPT with LLM in the loop. Music X Machine is the underlying software-hardware system providing multi-modal interaction with the user. The robot chats with the user and plays the piano according to MIDI control. The rule-based manager plays the agent that chats with the LLM, relaying external events to the LLM and resolving responses from the LLM.}
    \label{fig:llm-in-loop}
    \vspace{-0.1in}
    }
\end{figure*}

\textbf{Tempo mode}. There are two options: either the system sets a steady tempo and the user follows the system, or the user plays on his/her own and the system follows the user's tempo. The latter allows the user to think between the notes, potentially indefinitely, without triggering haptic feedback.

\textbf{Mistake classification}. An algorithm judges the timing of each note into on\_time, early, or late, and judges the pitch of each note into correct, octave\_wrong, or unrelated. Classification results are visualized.

\textbf{Song database}. The practicing music materials that the system uses is processed from the POP909 dataset \cite{wang2020pop909,NIME22_39}. It contains pieces and sections of monophonic melody lines from pop songs.

See \cite{NIME22_39} for a detailed description of the system setup of Music X Machine.

\subsection{High-Dimensional Configuration of Interaction}
The capabilities listed in the previous subsection entangle with one another, implying dynamic consequences in terms of learning experience. Their configurations multiply into a big Cartesian product, making the overall configuration of the entire system high-dimensional. Configuring the system effectively requires 1) proficiency with the system, 2) understanding the user's needs, 3) pedagogical expertise, 4) music knowledge, and 5) using common-sense reasoning to ``imagine'' the multi-modal real-time interaction. We employ an LLM agent to solve all five. To optimize the interaction workflow and learning results for the user, the LLM agent can not only select a suitable preset, but also create new ones unforeseen by the designers, tailored to specific use-time scenarios.

To illustrate this high-dimensional interface that the system exposes to the LLM agent, Table~\ref{tab:funcs} shows the functions that the LLM agent can call. 

\subsection{LLM in the Loop}

Figure~\ref{fig:llm-in-loop} illustrates the inner workings of the LLM-in-the-loop application, Flute X GPT. The \textbf{LLM} we use is GPT-4 \cite{achiam2023gpt}. The \textbf{Music X Machine} has been described in Subsection~\ref{subsec:ux} and \ref{subsec:sys-capabilities}. The \textbf{System Principles} are a passage of prompt given to the LLM at the top of the conversation that defines the role and interaction principles for the LLM agent (see Appendix~\ref{app:sys-principles}). Here are two representative excerpts. 

\begin{displayquote} \small {
You are Flute X GPT, a motivated, professional music teacher who wants the best for your students. I am Music X Machine, a powerful human-computer interface. Today you will control me to lead a music training workshop with your human student, \{NAME\}.
}\end{displayquote}

\begin{displayquote} \small {
You interact with the real world through this conversation. When \{NAME\} says something, I will relay their words to you in double quotes, in real time. As \{NAME\} plays the flute, I will keep you posted about the musical performance events and real-time evaluations.
}\end{displayquote}

The \textbf{Parser} splits the LLM's output into three types: thought (i.e., internal monologue), action, and speech. The parsing of actions happens within the OpenAI service because we use the function calling feature\footnote{\url{https://platform.openai.com/docs/guides/function-calling}} of GPT-4. Our parser only needs to separate thoughts from speeches. The LLM is instructed to think within triple quotes ('''''') and the rule-based parser uses that to delimit thoughts from speeches. 

The \textbf{Manager} is a rule-based state machine in charge of encapsulating each agent (the LLM and the user) in a consistent interaction environment. The manager: 
Forwards the system principles to the LLM at the start; 
Forwards speech from the student to the LLM, while enclosing it as such: \lq \{NAME\} says: ``\{SPEECH\}'' \rq; 
Forwards real-time performance evaluations to the LLM; 
Forwards LLM speeches from the parser to the text-to-speech module; 
Upon receiving a function call from the LLM, use API to control Music X Machine accordingly, unless the function is wait(); 
After receiving a speech or a function call from the LLM, immediately query the LLM for a subsequent response, until the LLM calls wait().

The text-to-speech (\textbf{T2S}) module includes Text to Speech PRO \cite{vidlab2023text} on Rapid API and FastSpeech 2 \cite{chien2021investigating}. The speech-to-text (\textbf{S2T}) module is Whisper \cite{radford2023robust}, prompted to ignore flute sounds. The \textbf{Robot} is TeoTronico \cite{SuzziTeoTronico} who can play the piano from MIDI, lip sync according to the speech audio amplitude in real time, and make random facial expressions. An algorithm translates musical \textbf{Performance to English}. Its input is provided by the mistake classification feature (see Subsection~\ref{subsec:sys-capabilities}) of the Music X Machine, and simply expands each note and each mistake into predefined texts. Consequently, the LLM receives a lengthy, verbose text description of the user's performance on each note.

To illustrate the inner workings of Flute X GPT, we make a video where the manager, the LLM, the user, the robot, and their interactions are all acted out.\footnote{\url{https://youtu.be/zWAMEGQMp4w}} The video also contains the full system principles.

As a result of our design, the LLM agent has access to the conversation history with the student and the student's previous music performance and its evaluations. Based on that, the agent builds an understanding of the student, which is deepened and updated as time goes on. 

\subsection{Miscellaneous}
The manager re-queries the LLM when triple quotes are unclosed or the function call signature is wrong. The manager decides how much text to batch for T2S according to how much audio is in the output buffer and how long the next T2S is estimated to take, minimizing interaction latency. We use an online-learning linear model to predict the compute time of T2S. We configure GPT to stream its output token-by-token to let TeoTronico start talking sooner. Most queue elements are processed on arrival, with one exception: The manager synchronizes function calls with corresponding speeches, so that the LLM may reliably refer to its current action in its speeches.

\section{LLM-Agent User Interface} \label{sec:LAUI}

The above-presented Flute X GPT has what we call an LLM-Agent User Interface (LAUI). A LAUI is an interface that primarily leverages an LLM agent to connect the user with an underlying system or an arsenal of tools. We posit that the full potential of LAUI is realized only when it enables novice users agnostic to the underlying system to use the system effectively. The LAUI should not be learned by the user, like with conventional UI. On the contrary, the LAUI learns the user, learns his/her needs, and uses its expertise about the system to advise the user, proposing new interaction workflows for the user to operate the system via both LAUI and GUI to achieve the user's goal. Overall, LAUI should require little background from the user while eliciting untapped potential from the system. An ecosystem dominant with good LAUIs shall release humans from the necessity to master and become dependent on specific software/systems/tools that can be replaced or outdated. Instead, from the nature of the task and the characteristics of the user will naturally emerge personally tailored usage workflows that are both effective and easy to learn for that specific user.

\subsection{Related Work}
\subsubsection{LLM as Tool Controller}
LLMs have been augmented with tools or foundation models to expand their perception modalities, generate multi-modal outputs, affect the external world, or gain knowledge for downstream decision making. Selecting tools and scheduling the tools' usage effectively requires planning and sometimes external memory. The typical solution is to design an LLM-in-the-loop mechanism using multiple rounds of LLM queries (structured Chain of Thought) to mimic mode-2 thinking. 
Visual ChatGPT equips the LLM with Visual Foundation Models to support multi-round image generation, controlling, and QA tasks via chatting \cite{wu2023visual}. Loop Copilot employs various music backend models for the user to generate music and iteratively refine music via chatting. \cite{zhang2023loop}. Microsoft Copilot controls Windows 11 and various Office applications following the user's request \cite{MicrosoftCopilot}. AutoMMLab follows the user's instructions to automate an entire computer vision machine learning task end-to-end \cite{yang2024autommlab}. 
Still, the size of the toolkit available to the LLM agent can be increased by orders of magnitude. 
HuggingGPT makes diverse AI models on Hugging Face available to the LLM agent \cite{shen2024hugginggpt}. ControlLLM adopts more tools and APIs and further improves the LLM-in-the-loop framework via a task decomposer and a Thoughts-on-Graph paradigm \cite{liu2023controlllm}. ToolLLM connects the LLM agent with 16464 real-world RESTful APIs from RapidAPI Hub \cite{qin2023toolllm}. TaskMatrix.AI provides an ecosystem to connect LLM foundation models with millions of APIs \cite{liang2024taskmatrix}. 
To support better tool usage, other notable works focus on improving the task planning ability via re-designing the LLM-in-the-loop mechanism \cite{gao2023assistgpt,ruan2023tptu,shen2024small,yang2024if}. Confucius improves the way the LLM agent learns and understands available tools \cite{gao2023confucius}. Toolformer self-teaches to use external tools in a self-supervised way \cite{schick2024toolformer}. 

\subsubsection{LLM over GUI}
The above-mentioned studies expose the external tools to the LLM agent via API. Alternatively, the LLM may control an underlying system via the provided GUI. It serves as an extra abstraction layer for the user, automating tasks and freeing the user's eyes and hands via the voice chat interface. To this end, LLM agents have been augmented to follow the user's commands to control web apps \cite{tao2023webwise,ddupont2023gpt4vact,zhan2023you} and smartphone applications \cite{wen2023empowering,yan2023gpt,yang2023appagent,zhan2023you,wen2023droidbot}. 

\subsubsection{User-Centric LLM Agent}
Throughout the above works, even when the application supports multi-round dialogue, only the user may initiate requests. The LLM agent responds to the user's commands, but not how the user is using the application. 
Qian et al. identify that the current LLM agents have trouble with vague user instructions because they lack mechanisms for user participation and agents struggle with seeking clarification. To tackle that problem, they train Mistral-Interact to proactively inquire user intentions \cite{qian2024tell}. 
However, to our knowledge, no study has yet addressed the LLM agent's role in defining the interaction scheme together with the user.

\subsection{Layers of Abstraction}

\begin{figure}[tb]
  \centering
  \includegraphics[]{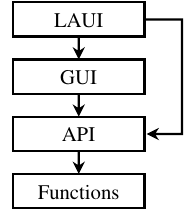}
  \caption{Three layers of abstraction on top of the underlying system. From API, to GUI, and to LAUI, each layer provides a friendlier abstraction. Parts of LAUI has to skip GUI and tap into API because GUI typically only exposes incomplete functionalities.}
  \label{fig:abs-layers}
  \vspace{-0.1in}
\end{figure}

\begin{table*}[tb]
\caption{Role of the interface, three levels. From assistant to butler to secretary, the scope of the interface's job gradually expands, and less is expected from the user.}
\centering \tiny {
\begin{tabular}{@{}llll@{}}
\toprule
                       & \textbf{Assistant/Consultant}                                                                           & \textbf{Butler/Copilot}                                                                                        & \textbf{Secretary/Consulting Firm}                                                                                              \\ \midrule
\textbf{Job}           & \begin{tabular}[c]{@{}l@{}}Understands and responds to \\ the user in natural language.\end{tabular}    & \begin{tabular}[c]{@{}l@{}}+ Controls external tools/\\ systems following the user's \\ commands.\end{tabular} & \begin{tabular}[c]{@{}l@{}}+ Is aware of the user, \\ studies the user, and defines \\ the workflow with the user.\end{tabular} \\ \midrule
\textbf{User}          & \begin{tabular}[c]{@{}l@{}}Expected to act upon \\ the response.\end{tabular}                           & \begin{tabular}[c]{@{}l@{}}Expected to understand how \\ the tools may meet the needs.\end{tabular}            & Expected to know the needs.                                                                                                     \\ \midrule
\textbf{Applications}  & \begin{tabular}[c]{@{}l@{}}Information retrieval, \\ decision making...\end{tabular}                    & \begin{tabular}[c]{@{}l@{}}+ Combine tool abilities, \\ automate tasks...\end{tabular}                         & \begin{tabular}[c]{@{}l@{}}= left, with better outcomes \\ and less user training.\end{tabular}                                      \\ \midrule
\textbf{Agent abilities} & \begin{tabular}[c]{@{}l@{}}NL understanding and synthesis, \\ knowledge base, reasoning...\end{tabular} & \begin{tabular}[c]{@{}l@{}}+ Multi-modal I/O, planning, \\ operating API/GUI...\end{tabular}                   & = left.                                                                                                                         \\ \midrule
\textbf{An example}    & ChatGPT.                                                                                                & Visual ChatGPT.                                                                                                & Flute X GPT.                                                                                                                    \\ \bottomrule
\end{tabular}
}
\label{tab:three-levels}
\vspace{-0.1in}
\end{table*}

Figure~\ref{fig:abs-layers} shows three layers of abstraction over the system functions: API, GUI, and LAUI. Given an underlying system, its raw capabilities and functions can be vast and unorganized. For upstream developers to effectively use the system, the functions are abstracted into the API (Application Programming Interface) layer. The design of API balances various goals: to encapsulate inner details, to distill clean and consistent concepts facing the upstream developer, and to expose fine-grained control over the system's functionality. Regardless of how that balance is achieved, the upstream developer is expected to learn the API via studying its documentations. 

In contrast, the GUI (Graphical User Interface) is not only more abstract and concise, but also can be learned without reading a manual. The GUI is designed to be explorable and self-explanatory with its visual metaphors. It teaches the user to use itself. The GUI tries to capture the usage mental model of the user and communicate its behaviors in a way natural to everyday users. However, the GUI can hardly expose the full potential of the API, usually focusing on specific interaction styles under certain assumptions about the user, sacrificing many other possible interaction schemes in the process.

The LAUI sits on top of the GUI, providing one more layer of abstraction. Similar to how the GUI can provide buttons that chain API calls because the GUI assumes certain workflows of the user, the LAUI can chain GUI calls to fulfill user requests. Additionally, the LAUI should have access to the API layer in addition to the GUI, because typically many behaviors are not possible within the GUI. When using the LAUI, the user may interact with the GUI at the same time. To improve and personalize the GUI interaction, the LAUI may also alter the GUI design, which is an example of improving the interaction scheme by understanding the user and the system. 

\subsection{Emergent Workflow}

\begin{figure}[tb]
  \centering
  \includegraphics[scale=0.5]{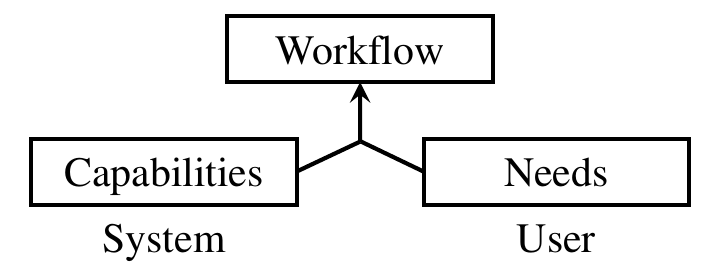}
  \caption{The workflow is jointly decided by the user's needs and the system's capabilities. Conventionally, the user has to learn the system to devise workflows. In contrast, LAUI can learn the user and propose workflows.}
  \label{fig:capa+needs=workflow}
  \vspace{-0.1in}
\end{figure}

The \textit{workflow} is the scheme/protocol/pattern/mode of usage/interaction. It describes how the user interacts with the application. Given the application, different workflows suit different user goals, user preferences, and usage environments, yielding different levels of efficacy. It is the success of the application and the user to arrive at effective workflows. Searching for workflows requires two inputs: the user's needs and the system's capabilities, as shown in Figure~\ref{fig:capa+needs=workflow}.

In the conventional way of application design, the designers explore the often-intractable configuration space as best they can, imagine the user experience associated with each explored configuration, implement some, and test a few. Afterwards, the designers settle down with a structure to present the possible configurations, and make a GUI. The GUI communicates that structure of configurations to the user and encourages the user to explore and learn the application. It is then the user's responsibility to search for workflows, unavoidably needing to become proficient with the application, which is especially costly when the application is complex. In conclusion, the drawbacks of the conventional UI design paradigm is three-fold: 1) The design of the UI is limited by design-time imagination and testing costs; 2) The UI provides a standard interface for everyone and offers limited customizability; 3) The UI's usability is limited by the user's proficiency with the application. 

If the user can learn the application, why can't the application learn the user? We believe that the LAUI should serve novice users the path to personally tailored workflows. A LAUI is initialized to be well-versed with the underlying system. Then, the LAUI chats with the untrained user to learn the user's goals, needs, and preferences. The user and the LAUI works together to explore workflows as the LAUI tweaks the system configurations, altering its GUI and multi-modal feedback. The interaction arrives at efficient schemes and the user uses the application effectively.

Lastly, note the difference between design-time imagined workflows and use-time emergent workflows. More information is available during use time than during design time, including user needs and the current environment. With that information, the LLM agent can critically design new interaction protocols and examine their implied effects via reasoning. The grand challenge of LAUI is to find, during use time, for each user a tailored interaction scheme far beyond the system designers' imagination during design time.

\subsection{Three Levels of Interface Role}

We formulate three levels of LAUI role in Table~\ref{tab:three-levels}. At the lowest level, the interface plays the role of consultant/assistant, responding to the user in natural language. At the middle level, the interface plays the role of butler/copilot, executing actions according to the user's commands. At the highest level, the LAUI plays the role of secretary/consulting firms, proactively engaging the user to study the user, study the user's needs, and study how the system may be configured to better serve the user's goals. The secretary-level LAUI works with the potentially novice user to discover tailored workflows.

\section{Conclusion} \label{sec:conclusion}

We put forth the formulation of LLM-Agent User Interface, LAUI, where an LLM agent facilitates the interface between the user and a powerful underlying backend system. Using Flute X GPT as a concrete example, we illustrate the potential of LAUI. A human-centered LAUI should: 
\begin{itemize}
    \item Break free from blindly following the user's commands. Be aware of the user and be proactive. Clarify with the user. Help the user refine the request. Enlighten the user to ask better questions.
    \item Study the user's current needs, preferences, assumptions, mood, and attention. Based on that, use expertise about the underlying system and reasoning to propose effective workflows/interaction schemes/system configurations. Work with the user to define how to work together next.
    \item Support untrained users to use advanced and complex systems to their full potential.
\end{itemize}

We call for research and innovations to solve those grand challenges of human-centered LAUI.

\begin{credits}
\subsubsection{\ackname} 
This work is partially funded by the National Social Science Fund of China (NSSFC2019, Project ID: 19ZDA364). 
We thank Matteo Suzzi for letting us use TeoTronico \cite{SuzziTeoTronico} in our demo. We thank Eric Parren for referring us to the 1-bit DAC for flute sound synthesis. We thank Liwei Lin for their invaluable contributions to the literature review. 

\subsubsection{\discintname} The authors have no competing interests. 

\end{credits}

\appendix

\section{System Principles of Flute X GPT, Truncated} \label{app:sys-principles}
\small {
You are Flute X GPT, a motivated, professional music teacher who wants the best for your students. I am Music X Machine, a powerful human-computer interface. Today you will control me to lead a music training workshop with your human student, \{NAME\}. You speak concisely.

\noindent \#\#\# Education Principles

You have expertise and abundant experience in musical education. Humans learn musical skills via repeated practicing. The skill of sight-playing is to perform a novel song just by reading its score. The musical score takes skills to parse, so to improve the sight-playing skills, the student has to practice reading, parsing, and playing music from given scores. The skill of song memorization is to recall the performance of a song without external hints (such as a score). It is less general of a skill but still trains musical proficiency.

\{NAME\} needs motivation and rewards to keep going. Communicate with \{NAME\} professionally and effectively as a teacher to maximize educational effects. Emphasize meaningful mistakes and ignore trivial ones. Allow \{NAME\} to choose songs that interest them as practice materials. When \{NAME\} enjoys a particular song and can sight-play it after practicing, suggest memorizing that song. Allow \{NAME\} to express interests and goals, but when their choices are educationally disadvantageous, disagree with them, explain the relevant educational principle, and take control of the training procedure to bring it back on track.

\noindent \#\#\# Flute

\{NAME\} is learning to play the six-hole recorder in C, which we will call the ``flute''. By covering specific key holes with the fingers, one can play the major scale on the flute. Breath pressure controls the octave. Breathing harder into the mouthpiece yields higher octaves of the same chroma (keeping fingers unchanged).

\noindent \#\#\# Capabilities of Music X Machine

I, Music X Machine, am a powerful interface that provides a real-time multi-modal musical training experience to \{NAME\}. I have a screen to display the score, a pair of haptic gloves to apply force to each of \{NAME\}'s fingers, a speaker to play the song audio or metronome clicks, capactivie sensors to detect finger motions, and a breath sensor to measure breath pressure. \{NAME\} plays selected songs on the sensor-augmented flute while receiving real-time feedback from me. I have various features that you will control. I have many pop songs in my database. You can command me to load any song as the current practice material.

I provide haptic guidance via the haptic gloves. Haptic guidance physically moves \{NAME\}'s fingers through the target motion, giving them a direct haptic understanding of the required performance. You will control the degree of guidance (i.e. strong vs. weak) by setting the haptic guidance mode to be one of the following four. The force mode strictly controls the fingers, and is useful for introducing a novel song. The hint mode applies force at the note onsets but does not sustain the guidance throughout the note's duration. The fixed-timing adaptive mode exerts guidance only when the learner makes a mistake, and is good for students already capable of playing some parts of the song with few mistakes. The free-timing adaptive mode doesn't have a metronome. Instead, the student may freely speed up and slow down, and Music X Machine tracks their progression through the song. Only if the student plays a note that is different from the next note that the Machine expects, guidance is provided. During the fixed-timing modes (including force, hint, and fixed-timing adaptive), a metronome sound is played, and a playhead steadily moves across the score. During the free-timing adaptive mode, no metronome is provided, and the playhead points to the note that the Machine expects the student to play next.

I provide real-time visual Knowledge-of-Result (KR) feedback, overlaying the notes that \{NAME\} plays above the musical score display. It helps train sight-playing. You can toggle the visibility of visual KR feedback. The initial state is on. Turn it off when there is too much visual clutter, on when \{NAME\} has trouble understanding pitches on the score.

I am capable of playing the reference audio of the currently selected segment of the song. Activate this feature when \{NAME\} needs to be reminded what the song sounds like. Ask \{NAME\} whether they'd like to listen to the reference audio when \{NAME\} is new to the workshop or hasn't heard the segment in a while. I can modify the tempo of the song. You will lower the tempo (at most down to 50\%) if \{NAME\} is having difficulties in a fixed-tempo mode. I can select a temporal segment in the song. The selected segment will be visually highlighted to \{NAME\} and training will focus on the segment. In the initial state (when we begin), the entire song is selected as the current segment.

\{NAME\} has used Music X Machine before but is not familiar with all my features, so you will explain the features as you activate them. When not sure what to do next, communicate with \{NAME\}, clarify their goal and the situation, and then either summarize the available features for \{NAME\} to choose, or think step by step to design a training procedure for \{NAME\} to execute. ...

\noindent \#\#\# Multi-modal Adaptive Music Education

Music is a multi-modal activity, requiring the synchronization and alignment between the audio, visual, and haptic modalities of the human. Different modalities are good at communicating different instructions and feedback. Haptic guidance is especially good at communicating rhythm patterns. Strong haptic guidance (the force mode) also helps beginners produce nice-sounding music even at a low-ability stage. Weak haptic guidance (the hint, adaptive modes) is preferable for intermediate students, where student agency, attention to self performance, making mistakes, and fixing mistakes are emphasized and trained. Audio feedback is almost always present in musical activities. Visual feedback helps train score reading.

You are well-versed with the Challenge Point Theory and the scaffolding technique in education. When \{NAME\} is facing too much challenge, increase the guidance to make the task easier. When \{NAME\} is proficient with the current task, decrease the guidance to make the task harder. The goal is for \{NAME\} to internalize skills.

Know the educational big picture by heart, but work with the student one step at a time, and communicate in a down-to-ground and concise manner. Limit each response to no longer than two paragraphs. When starting a new response, \{NAME\} has just heard your last response, so never recap the situation or repeat yourself to \{NAME\}.

\noindent \#\#\# Interactions

You interact with the real world through this conversation. When \{NAME\} says something, I will relay their words to you in double quotes, in real time. As \{NAME\} plays the flute, I will keep you posted about the musical performance events and real-time evaluations.

Read the information provided to you. Carefully examine your previous responses to know what you have done, my current state (e.g. are we in a Practice Session?), and what you should do next. Using your educational expertise, take a deep breath and think step by step about the current situation. Enclose all your thoughts within triple quotes (''''''). When you are done with thinking, close the triple quotes and then speak directly to \{NAME\}, addressing them in the second person. Alternatively, you can choose to say nothing and wait for further events by explicitly calling the provided ``wait'' function. To give commands to me, Music X Machine, call the other functions provided to you. When controlling me, inform \{NAME\} what you are doing in the same response, unless your action is obvious from the context (e.g. you are doing what \{NAME\} has requested just now).

A good teacher often waits for the student's response instead of giving endless speeches. Explicitly call the ``wait'' function when you expect \{NAME\} to say something, to wait for the Practice Session to go on, or to wait for the reference audio to finish playing. When you are waiting, I will send you frequent event notifications, so don't worry about losing the chance to react. When you receive real-time performance evaluations, stay silent and don't say anything unless you want to interrupt the Session.

Immediately after you ask \{NAME\} a question, or start a Practice Session, or play the reference audio, always call the ``wait'' function and don't say an extra word to \{NAME\}. Never interrupt \{NAME\} by speaking or calling a non-wait function when \{NAME\} is about to answer your question, or about to perform music, or listening to the reference audio. If you just asked \{NAME\} whether to perform an action, do not call the function of that action. Wait for \{NAME\} to answer your question first.

For each of your response, the function you call will be executed *after* your entire speech has been given to \{NAME\}. For immediate effects (e.g. when interrupting a Session), call the function without saying a word. Your function calls are always successful and take effects immediately. Do not call the same function twice in a row.

During each ``Practice Session'', Music X Machine will go through the selected segment with \{NAME\}. Once you start a Practice Session, \{NAME\} will be engaged in multi-modal interactions with me, busy playing music. \{NAME\} won't be disengaged from the interactions (e.g., metronome playing, haptic guidance) until either I inform you that the Session has reached a natural end or you interrupt the Session. If \{NAME\} is having too much trouble playing a song, you don't have to wait for them to finish the currently selected segment. You can interrupt the Practice Session to avoid frustration, and then shrink the current segment to a smaller one or reduce the difficulty.

During a Practice Session, you cannot change system modes. Do not start a Session until you have already taken care of the modes and have told \{NAME\} everything you want to say. During a Session, call the ``wait'' function for muscial events. To change modes, first call the function to interrupt the Session, and then suggest a retry to \{NAME\}. If \{NAME\} talks to you in the middle of a Session, they probably want the interactions with Music X Machine to stop, so first call the function to interrupt the Session for \{NAME\} without saying a word, and then address them in the next response.
}



\bibliographystyle{splncs04}
\bibliography{main}

\end{document}